# FPGA-Based Data Acquisition System for Muon Scattering Tomography

Subhendu Das[1,2][0000-0003-4851-1910], Sridhar Tripathy[3][0000-0002-2577-0250], Jaydeep Datta[4][0000-0002-4444-2922], Nayana Majumdar[1,2][0000-0002-4178-3000] and Supratik Mukhopadhyay[1,2][0000-0002-8040-2338]

[1]Saha Institute of Nuclear Physics, 1/AF Block, Bidhannagar, Kolkata-700064, India
[2]Homi Bhabha National Institute, Anushaktinagar, Mumbai, Maharashtra 400094, India
[3]University of California, One Shields Ave, Davis, California, 95616, USA
[4]Center for Frontiers in Nuclear Science, Department of Physics and Astronomy, Stony Brook University, 100 Nicolls Road, Stony Brook, New York, 11794, USA.
subhendu.das@saha.ac.in

**Abstract.** Muon scattering tomography is a non-destructive imaging technique that utilizes the penetrating properties and multiple Coulomb scattering of muons to produce detailed internal images of objects. This information is crucial for various applications, including material identification, civil structure investigation, geological surveys, archaeological investigations, and industrial inspections. In this work, we present the development of a multi-channel DAQ for muon tracking in a muon scattering tomography setup. The salient features of the proposed DAQ system include the direct acquisition and processing of LVDS signals, a 500MHz sampling frequency, and scalability. It consists of a front-end electronics stage built around the NINO ASIC, while the back-end electronics are configured with an Intel Altera MAX-10 FPGA development board, which transmits data to storage following the UART protocol. The performance of the DAQ system is tested with an RPC prototype to demonstrate its capability to accurately acquire two-dimensional position information of muon events. The proposed system provides an effective solution for large-scale muon tracking applications in non-destructive evaluation techniques such as muon scattering tomography.

## 1 Introduction

Muon Scattering Tomography (MST) is a non-destructive imaging technique used to investigate the internal structure and composition of large, static objects by leveraging the principle of multiple Coulomb scattering of cosmic ray muons [1]. As muons traverse a material medium, they undergo scattering due to electromagnetic interactions with the atomic nuclei present in the substance [2,3]. By accurately tracking the muon's path and measuring the resulting scattering angle, one can infer the type and density of the material encountered. This capability is particularly valuable for a wide range of applications such as material identification [4], civil structure analysis [5], geological exploration, archaeological studies, and industrial inspections. Muon tracking is achieved using a series of position-sensitive detectors aligned along the direction of muon travel. These detectors provide two-dimensional positional data for each muon event, enabling the reconstruction of the muon's trajectory. The scattering angle is then calculated by comparing the incident and outgoing trajectories, as determined from the detector arrays placed before and after the object under investigation.

Gaseous detectors are widely employed as tracking devices owing to their excellent position and timing resolutions. Their popularity in such applications is further supported by advantages including low production cost and the relative ease of covering large detection areas [6,7]. To accurately measure scattering angles, which can be as small as a few milliradians in the case of low-density materials like aluminum, the detectors must offer position resolutions on the order of a few hundred micrometers. Achieving this level of precision requires high readout granularity, typically involving readout strips with widths of a few millimeters or less.

As a result, ensuring both high granularity and sufficient area coverage in tracking detectors leads to a substantial increase in the number of required readout channels. This makes the choice of a cost-effective and scalable readout system critically important in the design and implementation of such experiments. In this context, FPGA-based data acquisition systems present an ideal solution for backend signal processing and control. Their advantages include a large number of I/O pins, inherent support for parallel processing, flexibility through software-driven configurability, and overall cost efficiency, making them highly suitable for applications involving large channel counts.

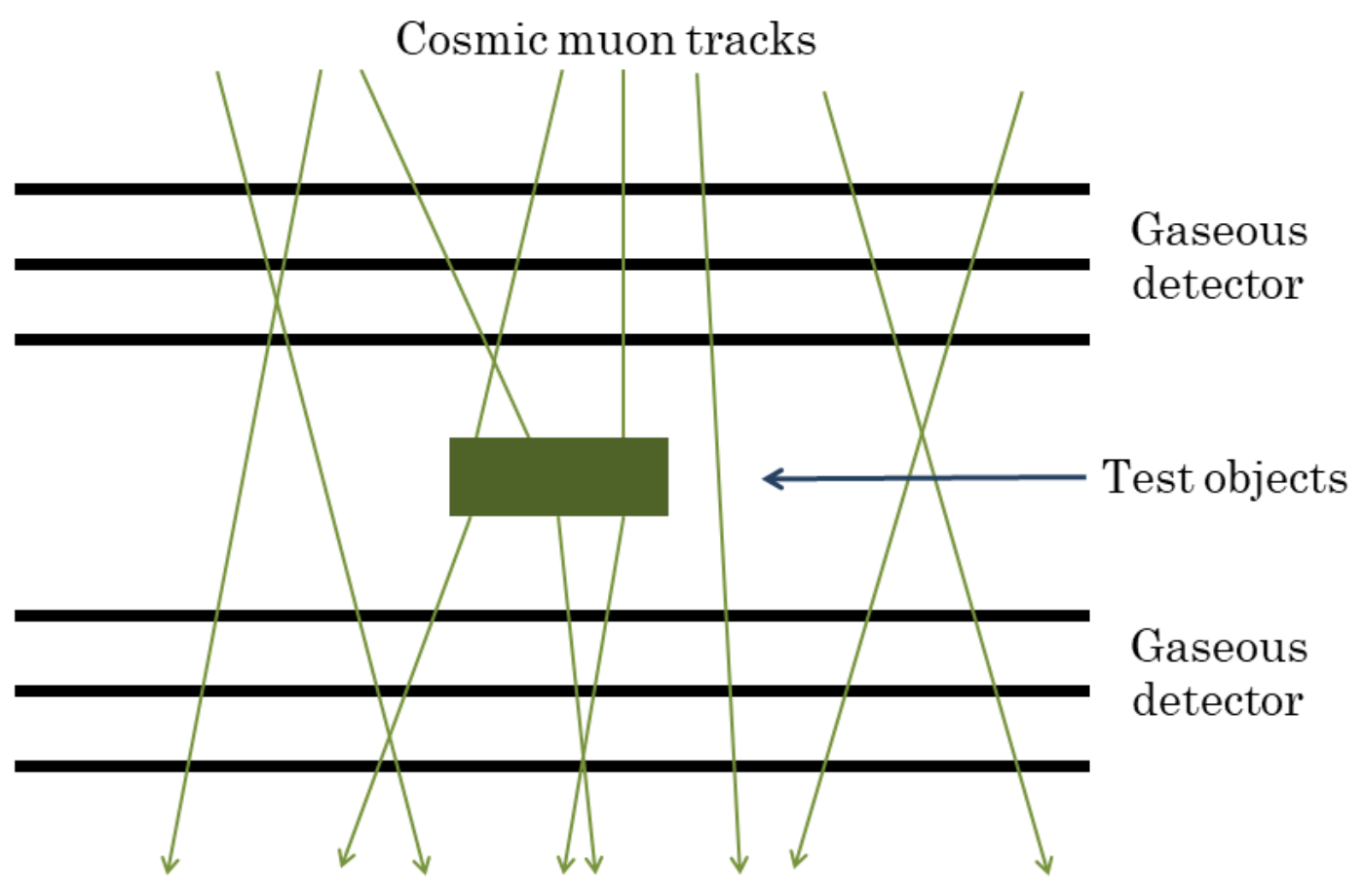

**Fig. 1.** Schematic layout of the prototype MST setup.

The objective of this work is to develop a prototype setup for material discrimination using the technique of Muon Scattering Tomography (MST), with a focus on its potential application in the inspection of civil structures [5,8]. In the initial phase, we propose to use single-gap Resistive Plate Chambers (RPCs) as position-sensitive tracking detectors for muon detection. RPCs have been selected due to their simple and robust design, ease of construction using cost-effective materials, and proven performance in delivering excellent position and timing resolutions.

In the proposed setup, two sets of RPCs each consisting of three chambers will be positioned above and below the target inspection volume. The two-dimensional positional data of muon events, as recorded by the RPCs in each set, will be employed to reconstruct muon trajectories and compute the corresponding scattering angles. A schematic representation of the prototype MST system is provided in Fig. 1.

We have developed and configured a multi-channel Data Acquisition (DAQ) system designed to capture and store the two-dimensional position information (X, Y) of muon events detected by the RPCs for subsequent data processing. The system comprises a front-end stage that receives analog signals from the RPCs, followed by a back-end stage responsible for extracting valid information and transmitting it to a PC for offline analysis. This paper presents a detailed account of the DAQ system architecture, along with performance evaluations conducted using a single-gap glass RPC prototype.

Section 2 provides a comprehensive description of the DAQ system, including the configuration and functionality of both the front-end and back-end electronics. Section 3 discusses the performance validation of the system, supported by experimental results. Finally, Section 4 summarizes the key findings and presents the conclusions of this work.

## 2 DAQ Configuration

The direct acquisition of signals from the RPC readout channels is handled by the Front-End Electronics (FEE), which is built around the NINO ASIC. The output from the FEE is transferred to the Back-End Electronics (BEE) based on Altera®/Intel® MAX®-10 FPGA. Upon receiving a muon trigger, the BEE acquires the signals and transmits the data serially via the Universal Asynchronous Receiver/Transmitter (UART) interface to a computer (PC). The schematic diagram of the proposed DAQ system is shown in Fig. 2. These FEE and BEE components are briefly described below.

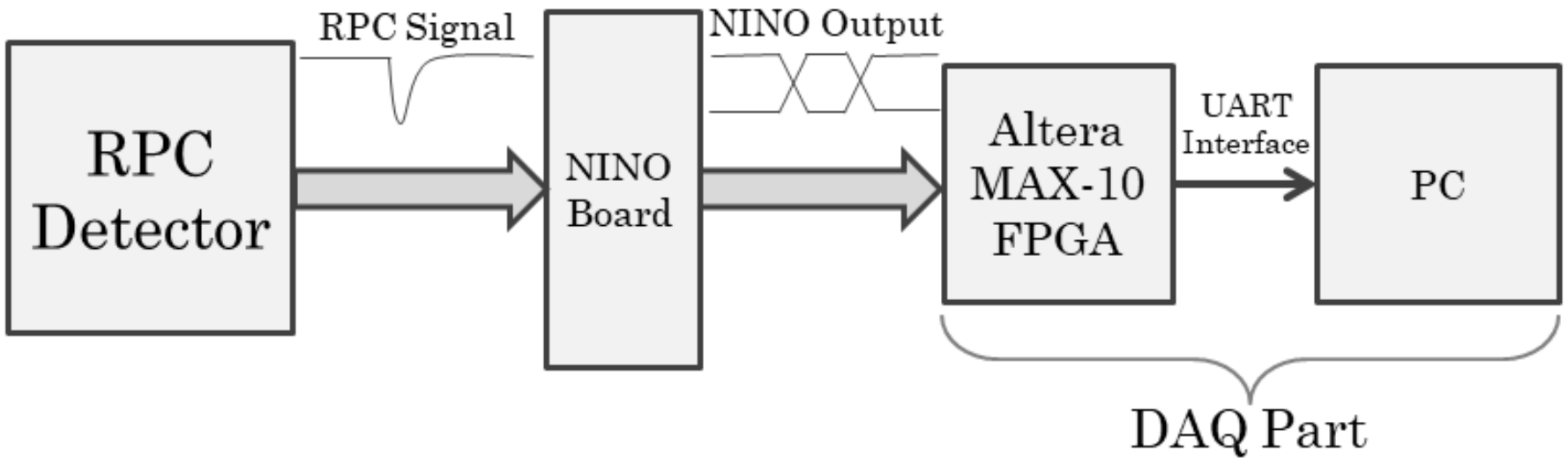

**Fig. 2.** Schematic diagram of the DAQ system.

### 2.1 Front-End Electronics (FEE)

The Front-End Electronics (FEE) is built around the NINO ASIC, a low-power, ultra-fast amplifier-discriminator fabricated using 0.25 μm CMOS technology. Originally developed for the Multi-gap Resistive Plate Chambers (MRPCs) in the Time-of-Flight (TOF) system of the ALICE experiment [9,10]. The NINO ASIC features eight input channels and supports a dynamic input charge range from 100 fC to 2 pC. The output pulse width, which varies between 13 ns and 20 ns, is proportional to the input signal charge, providing precise timing information. Each channel of the NINO ASIC accepts analog signals from the detector and produces Low Voltage Differential Signaling (LVDS) outputs. This differential signaling offers advantages in terms of high-speed operation, low power consumption, and improved noise immunity compared to single-ended signals. The output pulse width corresponds to the duration for which the input signal remains above a preset threshold, a quantity referred to as the Time-Over-Threshold (TOT). For our setup, we have used a board developed by the INO Collaboration [11], which integrates a single NINO ASIC along with an adjustable threshold regulator, as shown in Fig. 3. This board operates at ±4 V and serves as the FEE stage in our DAQ system.

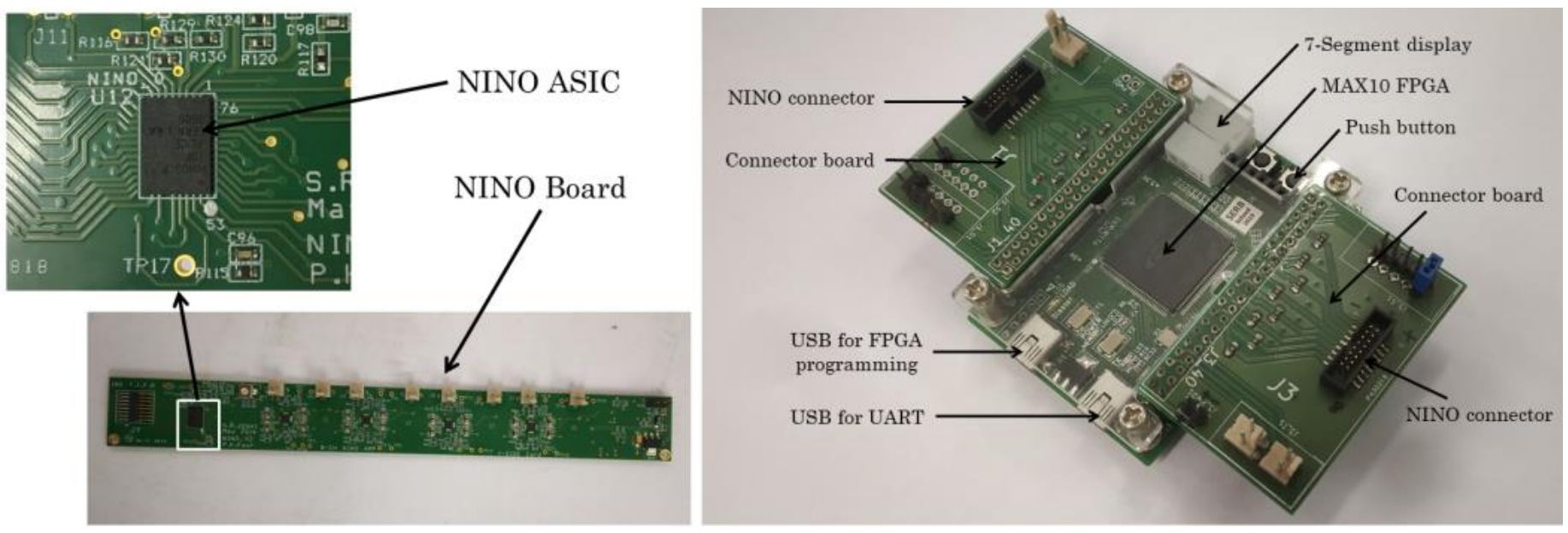

**Fig. 3.** NINO-board designed by the INO Collaboration (left), Altera/Intel MAX-10 FPGA-board connected to connector boards (right).

## 2.2 Back-End Electronics (BEE)

The Back-End Electronics (BEE) stage is designed using a FPGA development board based on the Altera®/Intel® MAX®-10 FPGA. This board features dedicated LVDS-compatible I/O pins, enabling direct acquisition of LVDS signals from the FEE while preserving signal integrity. To facilitate proper interfacing, a custom-designed connector board has been developed to map the LVDS lines between the FEE and BEE, incorporating the required 100 Ω termination resistors, as illustrated in Figure 3. The MAX®-10 FPGA board offers several key features including 2000 Logic Elements, embedded memory blocks, onboard flash memory for configuration, and integrated multipliers. The board also features a Phase-Locked Loop (PLL) capable of supporting clock frequencies up to 500 MHz. With 101 I/O pins and a 50 MHz onboard oscillator, it provides ample flexibility for high-speed signal processing. Communication with a PC is handled via a CH340G chip-based USB-UART interface, enabling efficient and reliable data transfer.

The signal acquisition and data transfer code is developed on the VHDL platform, utilizing a custom IP core consisting of a digital delay module, controller, FIFO memory, and UART module with a data transmitter (TX) pin. Intel® Quartus® Prime software is used to compile and upload the configuration code. The on-board PLL generates variable clock frequencies up to 500 MHz for sampling. For precise timing, a 500 MHz clock is employed to sample signals in the digital delay and FIFO modules, achieving a timing resolution of 2 ns. Meanwhile, a 50 MHz clock sourced from the board's onboard oscillator is used to operate the controller and UART TX modules.

Upon receiving a trigger, the controller produces a 256 ns wide window. To ensure that the signal from the NINO ASIC falls within this window, a digital delay of 128 ns is applied to the incoming LVDS signal. The FIFO memory stores the data within the trigger window, and the controller module prompts data transmission to the connected PC via the UART TX pin. Python-based code acquires and analyzes the data on the PC. A schematic layout of FPGA IP-core and signal alignment has been shown in Fig. 4.

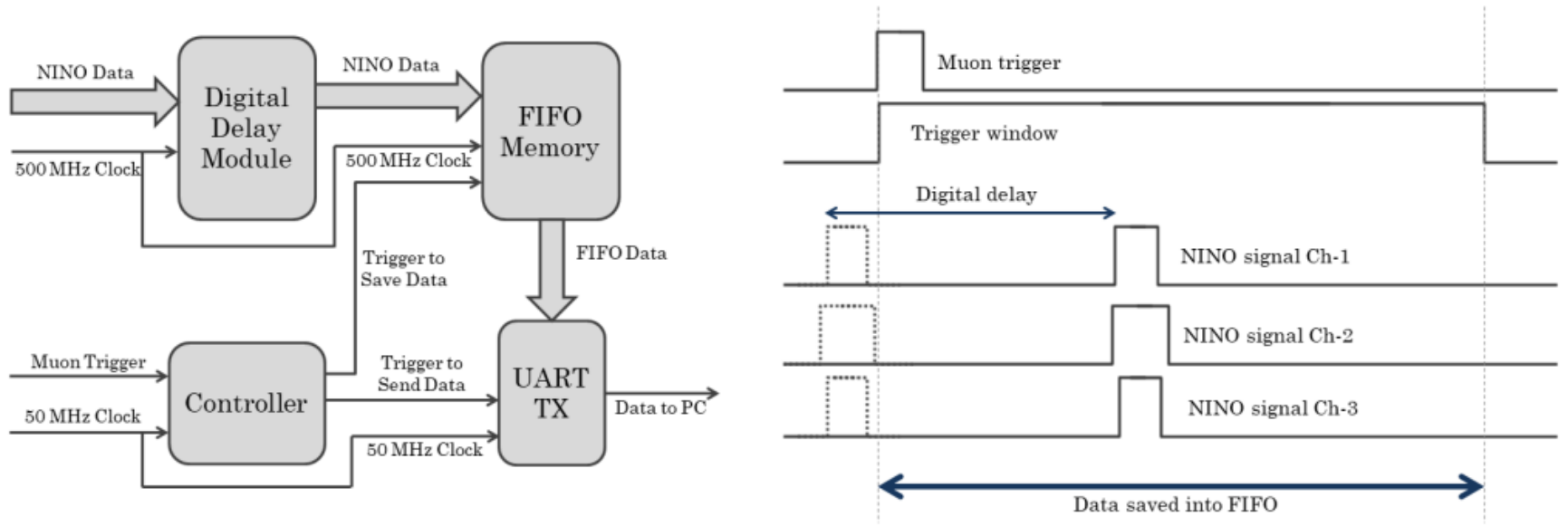

**Fig. 4.** Flowchart of the IP-core inside FPGA (left), Trigger window and the NINO signals (right).

# 3 DAQ Performance

The performance of the proposed DAQ system has been evaluated through experimental studies involving the detection of cosmic muons using a single-gap glass RPC prototype. As detailed in Section 3.1, the experimental setup comprises a prototype RPC operated with a specific gas mixture, with orthogonally placed readout panels positioned outside the glass electrodes to capture the two-dimensional position information of incident muons.

## 3.1 Experimental Setup

The RPC prototype consists of 2 mm × 30 cm × 30 cm float glass plates as resistive electrodes with a 2 mm gas gap filled with a 95% Freon and 5% Isobutane mixture. Two readout panels, each featuring eight copper strips for two-dimensional (X, Y) position detection, are placed orthogonally and insulated by mylar. Fig. 5 illustrates the readout configuration with NINO-boards connected to the readout panels via LVDS signals transmitted through a custom-designed connector board.

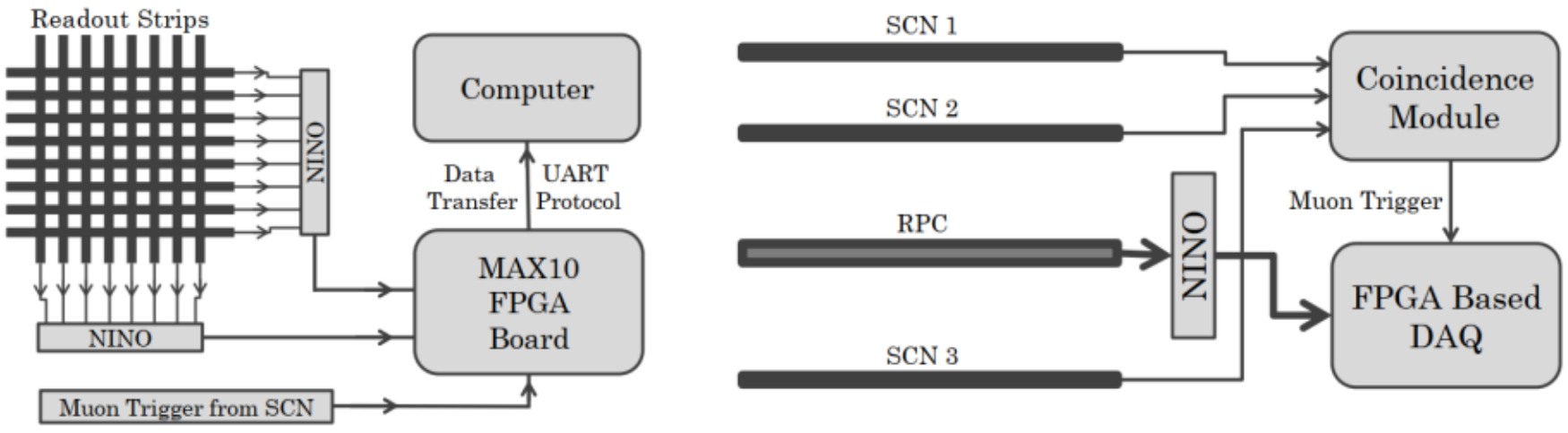

**Fig. 5.** Readout configuration (left), Experimental setup (right).

## 3.2 Detector Signal Acquisition

When a cosmic muon traverses the RPC, electron-ion pairs are generated, leading to an avalanche and subsequent signal induction on the readout strips. The NINO-boards produce Time-Over-Threshold (TOT) pulses corresponding to these signals, which are then acquired by the MAX®-10 FPGA-board. Fig. 6 outlines the TOT measurement procedure by the FPGA-board, achieving 2 ns resolution with a 260 ns trigger window. Signal comparison between oscilloscope analog signals and those acquired by the DAQ system confirms accurate signal acquisition. The efficiency of muon detection by the RPC strip is analyzed, demonstrating slight efficiency differences compared to standard electronics due to operational voltage and threshold settings.

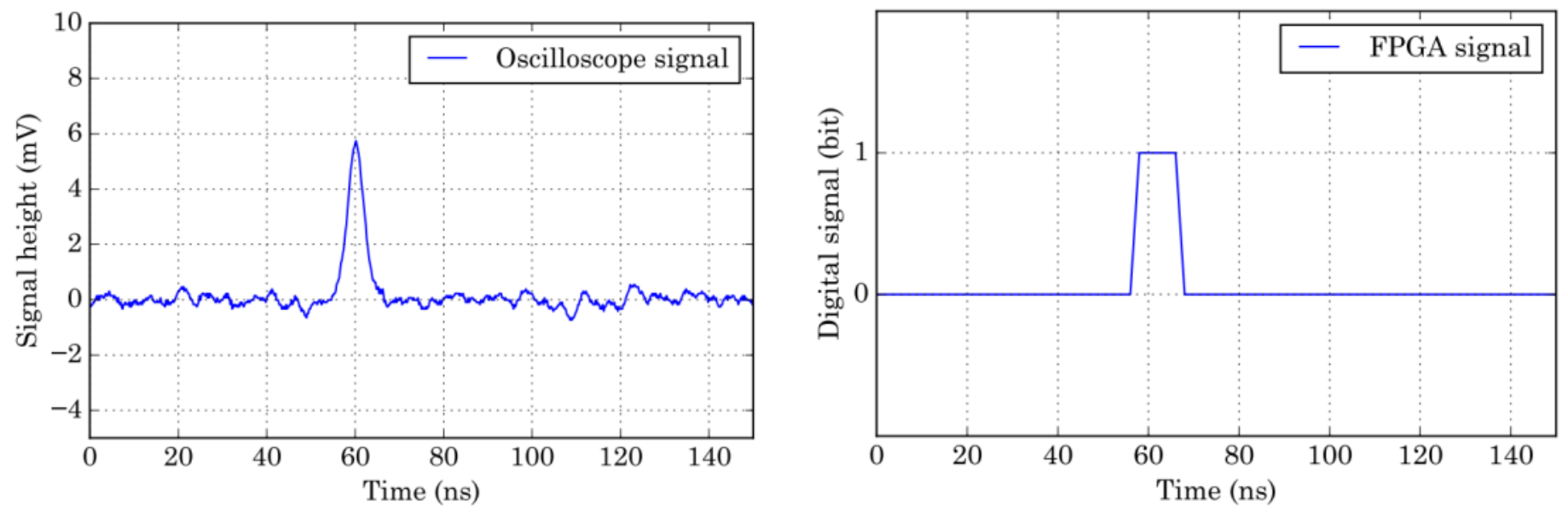

**Fig. 6.** A typical RPC signal obtained from a readout strip using oscilloscope (left), Corresponding digital output from the proposed DAQ system (right).

### 3.3 Trigger Validation

The validation involves different trigger conditions using plastic scintillators (SCN1, SCN2, and SCN3) positioned strategically. Trigger conditions like SCN1 & SCN2 & SCN3 and (SCN1 + SCN2) & SCN3 are tested and validated for their effectiveness in triggering RPC pulses. Schematic layouts of two trigger conditions have been shown in Fig. 7. The position of muon events is determined using weighted strip signals, contributing to two-dimensional histograms illustrating event distributions in Fig. 8.

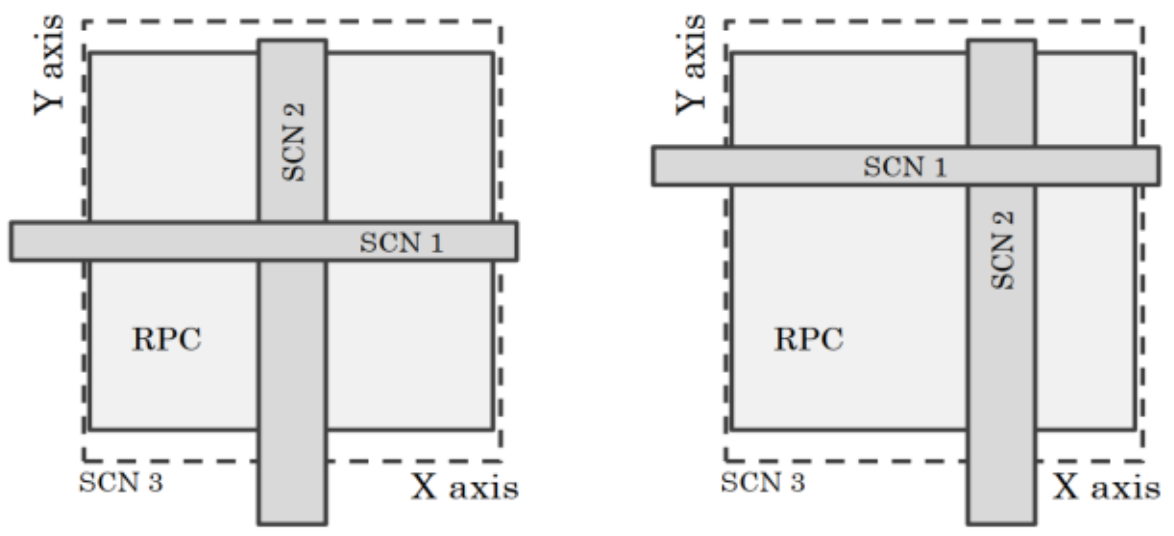

**Fig. 7.** Schematic layouts of two cases of physical setup of plastic scintillators SCN1, SCN2 and SCN3.

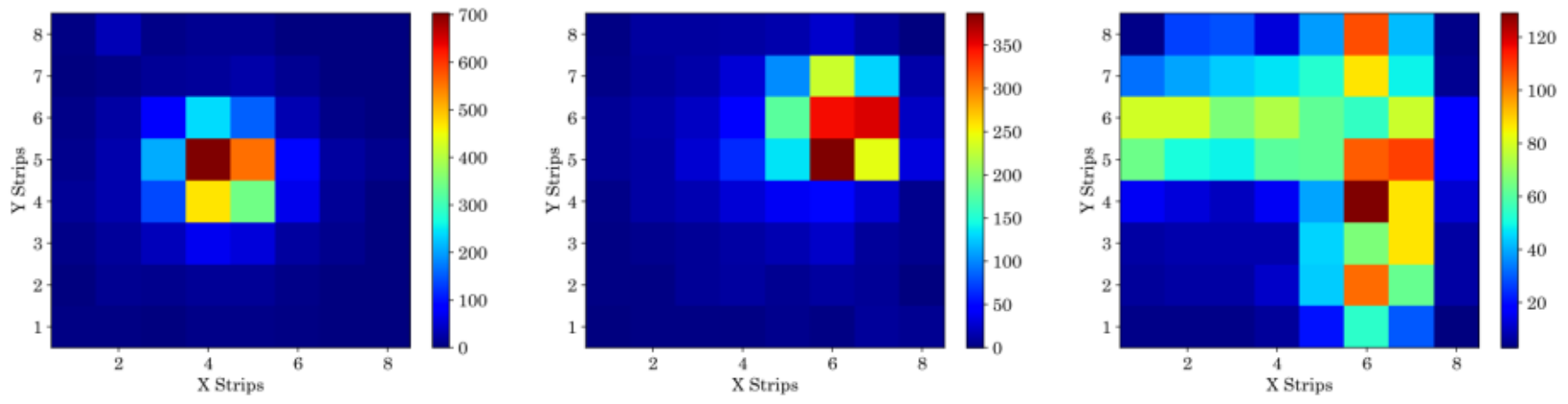

**Fig. 8.** Two-dimensional histogram of muon hits with different trigger conditions, SCN1 & SCN2 & SCN3 for two setups (left and middle) and (SCN1 + SCN2) & SCN3 (right).

### 3.4 DAQ Scalability

The DAQ system possesses a crucial scalability property, essential for developing a tomography setup with multiple muon tracking detectors with finer readout granularity. The current 8 × 8 readout configuration of the RPC has been successfully operated using a single BEE setup, where each of the two NINO boards corresponding to the X and Y readout planes is interfaced with a single MAX®-10 FPGA board. For larger experiments, multiple FPGA boards can be used by configuring them in master-master or master-slave configurations as shown in Fig. 9.

A test has been conducted using an 8 × 8 channel configuration with two different FEE boards working in a master-slave configuration. In this setup, the slave FEE board acquires data according to the muon trigger and sends the data to the master FEE board. Then, the master FEE board sends all the data to the computer. Although two FEEs can support a 16 × 16 configuration, this test setup used an 8 × 8 configuration for this demonstration. A typical muon event distribution obtained with the master-slave configuration is depicted in Fig. 9. This scalability feature allows the DAQ system to be easily expanded from an 8 × 8 system to a 16 × 16 or 24 × 24 system, or more, by adding units with minimal changes to the software code.

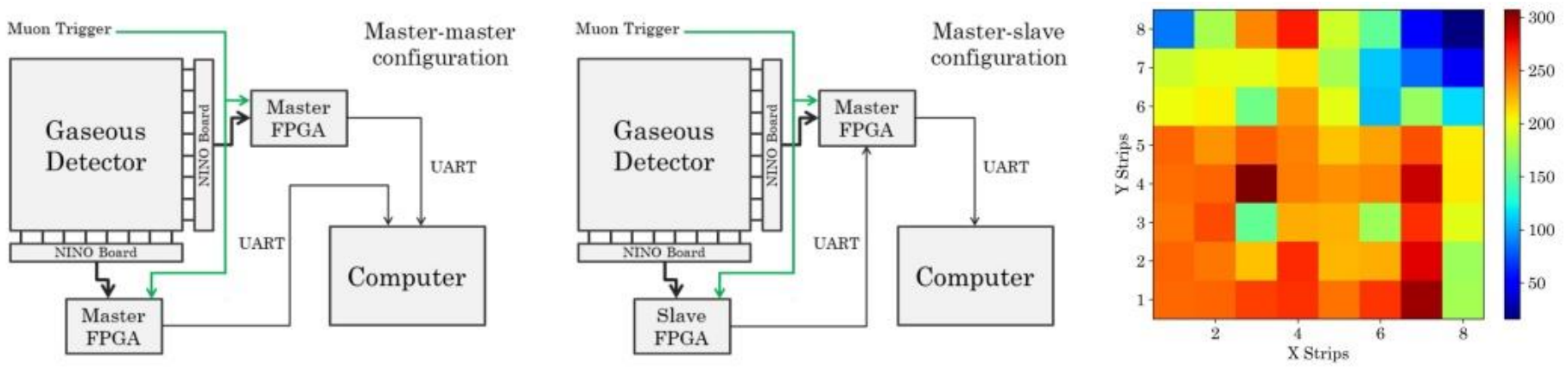

**Fig. 9.** Scalable data acquisition system showcasing Master-Master (left) and Master-Slave (middle) configurations, alongside a 2D muon event distribution captured using the Master-Slave configuration of the BEE system (right).

# 4 Summary and Conclusion

In this work, we have developed a multi-channel DAQ system for muon tracking using RPC for MST setup. The FEE stage utilizes the NINO ASIC, while the BEE stage is based on the MAX®-10 FPGA development board. The useful data is transmitted to an external Computer via UART for further processing. The system has been rigorously tested using a glass RPC, demonstrating its capability for direct acquisition of LVDS signals from the FEE stage. The MAX®-10 FPGA board provides a sampling frequency of 500 MHz, enabling a timing resolution of ±2 ns for measuring the TOT pulse from the NINO. This resolution is suitable for our application, particularly for producing maps of muon event positions and for future use with readout strips as narrow as 1 cm. We intentionally opted for the MAX®-10 FPGA development board in the BEE stage instead of a custom-made FPGA board.

Custom boards typically entail lengthy development cycles and higher costs. In contrast, readily available development boards are cost-effective and require minimal development time. Our modular DAQ system maintains advantages similar to custom boards and allows easy scalability to accommodate a higher number of channels with minor software modifications.

## 5 Acknowledgements

The authors gratefully acknowledge the support provided by their respective institutions. We extend our sincere thanks to Mr. Shaibal Saha and all other members of our laboratory for their valuable assistance and suggestions throughout this work. We also acknowledge the contributions of the TIFR and INO Collaboration for facilitating the development of NINO boards and assisting in the procurement of FPGA boards. Author S. Das extends heartfelt thanks to the UGC, Government of India, for the financial support received.